\documentclass[aip,
 ajp, amsmath,amssymb,
 reprint]{revtex4-2}

\usepackage{graphicx}
\usepackage{dcolumn}
\usepackage{bm}

\usepackage[utf8]{inputenc}
\usepackage[T1]{fontenc}
\usepackage{mathptmx}
\usepackage{etoolbox}

\makeatletter
\def\@email#1#2{%
 \endgroup
 \patchcmd{\titleblock@produce}
  {\frontmatter@RRAPformat}
  {\frontmatter@RRAPformat{\produce@RRAP{*#1\href{mailto:#2}{#2}}}\frontmatter@RRAPformat}
  {}{}
}%
\makeatother

\usepackage[percent]{overpic}

\usepackage{amsmath}

\makeatother
\usepackage{subcaption}

\usepackage{hyperref}    
\hypersetup{
    colorlinks=true,     
    linkcolor=blue,      
    urlcolor=blue,
    citecolor=blue
}

\makeatletter
\renewcommand*{\eqref}[1]{%
  \textup{\hyperref[#1]{\textcolor{blue}{(\ref*{#1})}}}%
}
\makeatother
\usepackage{float}
\newcommand{\Figref}[1]{\hyperref[#1]{\textcolor{blue}{Fig.~(\ref*{#1})}}}

\makeatletter
\newcommand{\iddots}{\mathinner{%
  \mkern1mu\raise1pt\hbox{.}\mkern2mu
  \raise4pt\hbox{.}\mkern2mu
  \raise7pt\hbox{.}\mkern1mu}}
\makeatother

\begin{document}

\title{Bloch's theorem: An operator-based derivation}
\author{Celal Sirin\textsuperscript{\hyperlink{references}{a)}} }
\affiliation{Department of Physics, Izmir Institute of Technology, 35430 Urla, Izmir, Turkey }
\date{\today}

\begin{abstract}
\noindent We present an operator-based derivation of Bloch's theorem. In this approach, we first construct an explicit operator form of the Hamiltonian by writing the potential part in terms of an infinite sum of the momentum-translation operators. Next, we prove that it commutes with the position-translation operator corresponding to a lattice translation, using the exponential re-ordering identity. We then build on Merzbacher's treatment of simultaneous diagonalization of commuting operators to reproduce Kittel's \textit{central equation} and also to derive a new formula for the simultaneous eigenket corresponding to a specific energy eigenvalue. Moreover, through the Born rule, we provide a precise account of the position and momentum probability distributions for Bloch states, and we examine the concept of crystal momentum. Finally, we apply our findings to the Kronig–Penney model, where we numerically depict the band structure and illustrate the delocalized nature of the single electron.

\vspace{0.5em}

\noindent (This article has been published in the \textit{American Journal of Physics}, \href{https://doi.org/10.1119/5.0310014}{https://doi.org/10.1119/5.0310014}.)

\end{abstract}

\maketitle

\section{Introduction}

In late 1927 at the University of Leipzig, Werner Heisenberg assigned his first graduate student, Felix Bloch, the task of applying the new quantum theory to the conductivity of metals as his doctoral thesis. Published in 1929, Bloch’s dissertation\textsuperscript{\hyperlink{references}{1}} on the topic became a cornerstone of condensed matter physics and was later recognized by being named Bloch’s theorem. Two years after Bloch's seminal work, in 1931, Ralph Kronig  and William Penney applied Bloch’s idea to a specific model in order to gain a clearer understanding of metals.\textsuperscript{\hyperlink{references}{2}} Nearly a century later, we present an alternative approach to these foundational concepts. We begin by offering an operator-based proof of Bloch’s theorem and then we continue by applying the idea to the Kronig–Penney model. 

There are two approaches typically used to derive Bloch's theorem: A differential-equation-theory approach and a condensed-matter-physics approach. The former is intrinsically tied to Floquet's theory.\textsuperscript{\hyperlink{references}{3}} The latter is the standard approach employed in most physics curricula.\textsuperscript{\hyperlink{references}{4,}}\textsuperscript{\hyperlink{references}{5,}}\textsuperscript{\hyperlink{references}{6}} These conventional approaches are formulated within a differential-equation framework, aiming primarily at solving Schrödinger’s wave-equation in position representation. Here, we present an operator-only approach to Bloch's theorem. By saying \textit{operator-only}, we mean that the entire derivation remains within the representation-free ket space. A brief comparison with the \textit{first proof} of Ashcroft\&Mermin\textsuperscript{\hyperlink{references}{4}} is provided in Appendix A.

In the following, after reviewing the preliminaries, we first construct an explicit operator form of the Hamiltonian. Using this Hamiltonian and employing the exponential re-ordering identity, we prove that the Hamiltonian operator commutes with the position-translation operator corresponding to a lattice translation. Then, the equation governing the structure of the simultaneous-eigenkets and the corresponding eigenvalues is derived by building on Merzbacher's treatment of commuting operators.\textsuperscript{\hyperlink{references}{7}} As will be seen, this equation ends up being the same as Kittel's \textit{central equation} in \hyperlink{references}{Ref. 6}, Eq. (8.41) in \hyperlink{references}{Ref. 4} and Eq. (3.37) in \hyperlink{references}{Ref. 5}, albeit derived from an entirely different approach. Beyond reproducing this classic result, we introduce a novel formula for the simultaneous eigenket associated with a specific energy eigenvalue. This simple explicit form enables a precise discussion of \textit{how} a single electron \textit{extends} through the entire lattice, while also giving students a concrete and accessible expression to explore. We conclude the paper by presenting a numerical approach on the Kronig-Penney model. Through this numerical analysis, we clearly illustrate the band structure as well as the position and momentum probability distributions, offering a hands-on demonstration for students.

\section{Preliminaries}

As a starting point, we review several essential ideas from non-relativistic quantum mechanics. The notation employed here essentially follows that of standard physics curricula,\textsuperscript{\hyperlink{references}{8}} with the exception of a minor pedagogic modification: We place hats onto the operators to avoid any potential misinterpretation. Let us begin with the notation for the position eigenstate:

\begin{equation}
\hat{x} \big|x'\big> = x'\big|x'\big>
\qquad
x' \in \mathbb{R}.
\label{position}
\end{equation}

\noindent Here, $\hat{x}$ is the position operator and $x'$ is the position eigenvalue corresponding to the position eigenket $\big|x'\big>$. We have similarly for the momentum eigenstate that:

\begin{equation}
\hat{p}_x \big|p_x'\big> = p_x'\big|p_x'\big>
\qquad
p_x' \in \mathbb{R}.
\label{momentum}
\end{equation}

\noindent Here, $\hat{p}_x$ is the momentum operator and $p_x'$ is the momentum eigenvalue corresponding to the momentum eigenket $\big|p_x'\big>$. Having introduced the fundamental observables, we now proceed by presenting an important operator identity (see Appendix B):

\begin{equation}
     e^{\hat{A}} \hat{C} e^{-\hat{A}} = \hat{C} + [\hat{A}, \hat{C}] + \frac{1}{2!} \big[ \hat{A}, [\hat{A}, \hat{C}] \big] + \dots 
\label{opidentityb}
\end{equation}

\noindent This is known as the Hadamard lemma (or, the Baker-Hausdorff lemma). Here, the commutator is defined as \mbox{$[\hat{A}, \hat{C}] = \hat{A} \hat{C} -  \hat{C} \hat{A}$}, and dots stand for the higher order nested commutators. This identity can be used to derive the position and momentum translation operators. To this end, let us use Eq.~\eqref{opidentityb} to evaluate the following product:

\begin{equation}
\exp( i \frac{\hat{p}_x}{\hbar} \Delta x')  \cdot \hat{x} \cdot \exp( -i \frac{\hat{p}_x}{\hbar} \Delta x') = \hat{x} + \Delta x' \cdot \hat{I} .
\label{eqn4}
\end{equation}

\noindent Here, we used the fact that $[\hat{x}, \hat{p}_x] = i \hbar \cdot \hat{I}$, where $\hat{I}$ denotes the identity operator, $i$ is the imaginary unit, and $\hbar$ is the reduced Planck constant. Multiplying Eq.~\eqref{eqn4} by $\big|x'\big>$ from the right, and then multiplying it by \mbox{$\exp(-i (\hat{p}_x/\hbar) \Delta x')$} from the left, we obtain:

\begin{equation}
    \hat{x} \cdot \Big( \exp( -i \frac{\hat{p}_x}{\hbar} \Delta x') \big|x'\big>      \Big) = (x' + \Delta x') \cdot \Big( \exp( -i \frac{\hat{p}_x}{\hbar} \Delta x') \big|x'\big> \Big).
    \label{eqn5}
\end{equation}

\noindent By comparing this with Eq.~\eqref{position}, we can identify \mbox{$\exp( -i \frac{\hat{p}_x}{\hbar} \Delta x') \big|x'\big> = \big|x' + \Delta x'\big> $}. Consequently, we call this exponential the \textit{position-translation operator} and denote it as:

\begin{equation}
\begin{aligned}
\hat{T}(\Delta x') = \exp\bigg(-i\cdot \frac{\hat{p}_x}{\hbar}\cdot \Delta x'\bigg)
\end{aligned}
\label{eqn6}
\end{equation}

\noindent such that

\begin{equation}
\begin{aligned}
    \hat{T}(\Delta x')  \big|x'\big> &= \big|x' + \Delta x' \big> \qquad \Delta x' \in \mathbb{R}.
\end{aligned}
\label{eqn7}
\end{equation}

\noindent Furthermore, by following a similar line of reasoning, one can derive the \textit{momentum-translation operator},

\begin{equation}
\begin{aligned}
\hat{B}(\Delta p_x') = \exp\bigg(+i\cdot \frac{\hat{x}}{\hbar}\cdot \Delta p_x'\bigg),
\end{aligned}
\label{eqn8}
\end{equation}

\noindent such that

\begin{equation}
\begin{aligned}
    \hat{B}(\Delta p_x')  \big|p_x'\big> &= \big|p_x' + \Delta p_x' \big> \qquad \Delta p_x' \in \mathbb{R}.
\end{aligned}
\label{eqn9}
\end{equation}

\noindent With the notation for the fundamental observables and the translation of their eigenkets in place, we now ask whether the momentum and position translation operators commute. To address this question, we are going to derive another important result: the exponential re-ordering identity. For that, we consider the product \mbox{$\exp(\xi \hat{x}) \cdot \exp(\zeta \hat{p}_x) \cdot \exp(-\xi \hat{x})$}, and use the braiding identity (see Appendix B) in the first line:

\begin{equation} \label{eqn10}
\begin{aligned}
e^{\xi \hat{x}} \cdot e^{\zeta \hat{p}_x} \cdot e^{-\xi \hat{x}} &= \exp(e^{\xi \hat{x}} \cdot \zeta \hat{p}_x \cdot e^{-\xi \hat{x}}) \\
&= \exp ( \zeta \hat{p}_x + i \xi \zeta \hbar \cdot \hat{I}) \\
&= e^{i \xi \zeta \hbar \cdot \hat{I}} \cdot e^{\zeta \hat{p}_x}.
\end{aligned}
\end{equation}

\noindent In the second line, we use Eq.~\eqref{opidentityb} inside the exponential, and in the third we simply note $[\hat{p}_x, \hat{I}] = 0$. Now, let us multiply Eq.~\eqref{eqn10} by $\exp (\xi \hat{x})$ from the right:

\begin{equation}
    e^{\xi \hat{x}} \cdot e^{\zeta \hat{p}_x} = e^{i \xi \zeta \hbar \cdot \hat{I}} \cdot e^{ \zeta \hat{p}_x} \cdot e^{ \xi \hat{x}}.
    \label{eqn11}
\end{equation}

\noindent This is known as the exponential re-ordering identity. By letting $\xi = i\Delta p_x' / \hbar$ and $ \zeta = -i\Delta x' / \hbar$, we see that, in general, $[\hat{T}(\Delta x'), \hat{B}(\Delta p_x')] \neq 0$. However, we note that they do commute if the product, $\Delta x' \Delta p_x'$, is an integer multiple of $ 2 \pi \hbar$. We will encounter the realization of this exceptional case in the next section, which then forms the very foundation of this operator-based approach.

\section{Bloch's Theorem}

We consider a periodic potential energy with a period of $a$: $V(x' + a) = V(x')$. Such a function can be written in terms of its Fourier series expansion:

\begin{subequations} \label{eqn12}
\begin{align}
    V(x') &= \sum_{\ell=-\infty}^{\infty} V_{\ell} \cdot \exp \Big( i \cdot \frac{2\pi \ell}{a} \cdot x' \Big), \label{eqn12a} \\[10pt]
    V_{\ell} &= \frac{1}{a} \cdot \int_{-a/2}^{a/2}  V(x') \cdot \exp \Big( -i \cdot \frac{2\pi \ell}{a} \cdot x' \Big) dx'. \label{eqn12b}
\end{align}
\end{subequations}

\noindent Here, Eq.~\eqref{eqn12a} is the eigenvalue of the potential energy operator, $V(\hat{x})$, such that:

\begin{equation}
V(\hat{x}) \big| x' \big> = V(x') \big| x' \big>.
\label{eqn13}
\end{equation}

\noindent Now, instead of plugging the eigenvalue, $V(x')$, into the wave-equation, we aim to deduce the explicit operator form of it, $V(\hat{x})$. We find that it can be written simply by replacing the eigenvalue, $x'$, with the operator, $\hat{x}$, in Eq.~\eqref{eqn12a}:

\begin{equation}
V(\hat{x}) = \sum_{\ell=-\infty}^{\infty} V_{\ell} \cdot \exp \bigg( i \cdot \frac{2\pi \ell}{a} \cdot \hat{x} \bigg).
\label{eqn14}
\end{equation}

\noindent This is the key conceptual leap in our line of reasoning. Equation \eqref{eqn14} can be verified by plugging it into Eq.~\eqref{eqn13}. To proceed, we now note that the summand in Eq.~\eqref{eqn14} is nothing but the momentum-translation operator introduced in Eq.~\eqref{eqn8}, with $ \Delta p_x' = 2 \pi \hbar \ell / a$. Therefore, we can write the full Hamiltonian operator as:

\begin{equation}
\hat{H} = \frac{\hat{p}_x^2}{2m} +  \sum_{\ell=-\infty}^{\infty} V_{\ell} \cdot \hat{B}\Big(\frac{2\pi \hbar \ell}{a}\Big),
\label{eqn15}
\end{equation}

\noindent such that

\begin{equation}
\hat{H} \big|H'\big> = H' \big|H'\big>.
\label{eqn16}
\end{equation}

\noindent Having recast the problem in the representation-free ket space, now, our task is to find the energy eigenvalue, $H'$, and also the corresponding energy eigenket, $\big|H'\big>$. To do that, we are going to first prove: $[\hat{H}, \hat{T}(a)] = 0$. Then, we are going to build on Merzbacher's treatment of commuting operators. As will be seen, the placeholder symbol $H'$ in Eq.~\eqref{eqn16} will later be replaced and augmented with additional labels. Let us start by considering the commutator:

\begin{equation}
\begin{aligned}
\Big[ \hat{H}, \hat{T}(a) \Big] = 
&\quad \Bigg[ \frac{\hat{p}_x^2}{2m}, \exp\big( -i \cdot \frac{\hat{p}_x}{\hbar} \cdot a \big) \Bigg] \\
&\quad + \Bigg[ \sum_{\ell=-\infty}^{\infty} V_\ell \cdot \exp \bigg( \frac{i2\pi \ell}{a}  \hat{x} \bigg), \exp \bigg( \frac{-i a}{\hbar} \hat{p}_x \bigg) \Bigg].
\end{aligned}
\label{eqn17}
\end{equation}

\noindent The first commutator vanishes trivially, as can be seen by plugging in the operator-series for the exponential. We are left with the second term:

\begin{equation}
\Big[ \hat{H}, \hat{T}(a) \Big] = \sum_{\ell=-\infty}^{\infty} V_\ell \Bigg[ \exp \bigg( \frac{i2\pi \ell}{a}  \hat{x} \bigg) , \exp \bigg( \frac{-i a}{\hbar} \hat{p}_x \bigg) \Bigg].
\label{eqn18}
\end{equation}

\noindent Using Eq.~\eqref{eqn11} and noting that $\exp( i 2 \pi \ell \cdot \hat{I} ) = \hat{I} $ for all $\ell \in \mathbb{Z} $, we obtain:

\begin{equation}
\Big[ \hat{H}, \hat{T}(a) \Big] = 0.
\label{eqn19}
\end{equation}

\noindent Next, we find the eigenkets and the eigenvalues of $\hat{T}(a)$:

\begin{equation}
\hat{T}(a) \bigg|   p_x' +  \frac{2 \pi \hbar j}{a}    \bigg> = \exp \big(-i \cdot \frac{p_x'}{\hbar} \cdot a \big) \bigg|p_x' + \frac{2 \pi \hbar j}{a} \bigg>.
\label{eqn20}
\end{equation}

\noindent This can be shown by recalling Eqs.~\eqref{momentum} and \eqref{eqn6}, and plugging in the operator-series for the exponential. We see that each eigenvalue of $\hat{T}(a)$ is infinitely degenerate. That is, $ \big|p_x' + 2 \pi \hbar j / a\big> $, for all integers $j$, is an eigenket of $\hat{T}(a)$ with the same eigenvalue. Now, keeping this in mind, let us act on Eq.~\eqref{eqn20} by $\hat{H}$ from the left and use Eq.~\eqref{eqn19} to swap the ordering in the left-hand-side. We have:

\begin{equation}
\begin{aligned}
\hat{T}(a) \cdot \Big(  \hat{H} \big|p_x' + 2 \pi \hbar j / a\big> \Big)= \\ 
 \exp \big(-i \cdot \frac{p_x'}{\hbar} \cdot a \big) \cdot \Big( \hat{H}\big|p_x' + 2 \pi \hbar j / a\big> \Big).
\end{aligned}
\label{eqn21}
\end{equation}

\noindent This equation tells us that the ket, $\hat{H} \big|p_x' + 2 \pi \hbar j / a\big>$, is also an eigenket of $\hat{T}(a)$ with the same eigenvalue. Therefore, we can write it as a sum in terms of the degenerate set of eigenkets as follows:

\begin{equation}
\hat{H} \big|p_x' + 2 \pi \hbar j / a\big> = \sum_{n=-\infty}^{\infty} C_{j,n} \big|p_x' + 2 \pi \hbar n / a\big>.
\label{eqn22}
\end{equation}

\noindent This result will prove useful shortly. Now, we choose the following to be the simultaneous-eigenket of both $\hat{H}$ and $\hat{T}(a)$:

\begin{equation}
\big| simul \big>_{p_x', H'} = \sum_{k=-\infty}^{\infty} D_{H',k} \big|p_x' + 2 \pi \hbar k / a\big>.
\label{eqn23}
\end{equation}

\noindent This is already an eigenket of $\hat{T}(a)$ with the eigenvalue of $\exp (-i p_x'  a / \hbar )$. Note how the subscript, $p_x'$, keeps track of this eigenvalue (see Appendix A). Next, we require that Eq.~\eqref{eqn23} is also an eigenket of $\hat{H}$:

\begin{equation}
\hat{H}\big| simul \big>_{p_x', H'} = H' \big| simul \big>_{p_x', H'}.
\label{eqn24}
\end{equation}

\noindent We wish to find the condition that makes this equality hold. For that, let us evaluate the left-hand-side (LHS) and the RHS separately, and then equate. Considering the LHS of Eq.~\eqref{eqn24} and plugging into Eq.~\eqref{eqn23}, and also using Eq.~\eqref{eqn22}, we obtain:

\begin{equation}
\begin{aligned}
LHS_{(24)} &= \hat{H} \big| simul \big>_{p_x', H'} \\
    &= \sum_{k=-\infty}^{\infty}
       \sum_{n=-\infty}^{\infty} D_{H',k} 
       C_{k,n} \big| p_x' + 2\pi\hbar n / a \big>.
\end{aligned}
\label{eqn25}
\end{equation}

\noindent Now, plugging Eq.~\eqref{eqn23} into the RHS of Eq.~\eqref{eqn24} and then equating it to Eq.~\eqref{eqn25}, we obtain:

\begin{equation}
\begin{aligned}
     \sum_{k=-\infty}^{\infty} 
       \sum_{n=-\infty}^{\infty} D_{H',k} 
       C_{k,n} \bigl| p_x' + 2\pi\hbar n / a \bigr> \\
    = \sum_{k=-\infty}^{\infty} H' D_{H',k} \big|p_x' + 2 \pi \hbar k / a\big>. \\
\end{aligned}
\label{eqn26}
\end{equation}

\noindent To proceed, we now recall that the momentum eigenkets are known to form a linearly-independent set. Therefore, for a certain $q$ in the $k$-sum/$n$-sum in Eq.~\eqref{eqn26}, the coefficients must be the same: \mbox{$\sum_{k=-\infty}^{\infty} D_{H',k} C_{k,q} = H' D_{H',q}$}. This can be compactly rewritten as:

\begin{equation}
\begin{aligned} 
       \sum_{k=-\infty}^{\infty}  
       ( C_{k,q} - H' \delta_{k,q} )  D_{H',k} = 0.
\end{aligned}
\label{eqn27}
\end{equation}

\noindent Here $\delta_{k,q}$ denotes the Kronecker delta. Compare this with Eq. (8.64) in \hyperlink{references}{Ref. 7}. From Eq.~\eqref{eqn19} to Eq.~\eqref{eqn27}, we have closely followed Merzbacher's treatment. From this point onward, however, our work builds upon it. In particular, we now employ our key insight, Eq.~\eqref{eqn15}, to determine the coefficients, ${C_{k,q}}$, in Eq.~\eqref{eqn27}. For that, let us revisit Eq.~\eqref{eqn22}. There, replace $j$ with $k$ because we seek $C_{k,q}$, and similarly evaluate the LHS of Eq.~\eqref{eqn22} using Eqs.~\eqref{eqn15}, \eqref{momentum} and \eqref{eqn9}:

\begin{equation}
\begin{aligned}
LHS_{(22)} &= \hat{H} \big|p_x' + 2 \pi \hbar k / a\big> \\
    &= \Bigg( \frac{\hat{p}_x^2}{2m} + \sum_{\ell=-\infty}^{\infty} V_{\ell} \cdot \hat{B}\Big(\frac{2\pi \hbar \ell}{a}\Big) \Bigg) \big|p_x' + 2 \pi \hbar k / a\big>  \\
    &= \frac{(p_x' + 2 \pi \hbar k / a)^2}{2m} \big|p_x' + 2 \pi \hbar k / a\big>  \\
    &\quad  + \sum_{\ell=-\infty}^{\infty} V_{\ell} \big|p_x' + 2 \pi \hbar (k+\ell) / a\big>.
\end{aligned}
\label{eqn28}
\end{equation}

\noindent Now, substituting Eq.~\eqref{eqn28} for the LHS of Eq.~\eqref{eqn22}, and similarly equating the coefficients of the momentum-eigenkets when we have $ \big|p_x' + 2 \pi \hbar q / a\big> $ in the sum, we obtain:

\begin{equation}
\begin{aligned}
     \frac{(p_x' + 2 \pi \hbar k / a)^2}{2m} \delta_{k,q} + V_{q-k}  
    =  C_{k,q}.  \\
\end{aligned}
\label{eqn29}
\end{equation}

\noindent To proceed further, for notational simplicity, let us rename the first coefficient above, as: \mbox{$\scalebox{1}{$\kappa$}_{p_x',k} \equiv (p_x' + 2 \pi \hbar k / a)^2 / 2m $}. Now, with this concise definition, we plug Eq.~\eqref{eqn29} into Eq.~\eqref{eqn27}:

\begin{equation}
\begin{aligned} 
       \sum_{k=-\infty}^{\infty}  
       \Bigg( \bigg( \scalebox{1.0}{$\kappa$}_{p_x', k} - H' \bigg) \delta_{k,q} + V_{q-k}  \Bigg)  D_{H',k} = 0.
\end{aligned}
\label{eqn30}
\end{equation}

\noindent This is our central result. By fixing $q$ to an integer and varying $k$, then fixing $q$ to another integer and varying $k$, and repeating this procedure; it can be seen that Eq.~\eqref{eqn30} is actually describing the following matrix equation:

\begin{equation}
\resizebox{\linewidth}{!}{$
\renewcommand{\arraystretch}{2.0} 
\arraycolsep=1.5pt 
\begin{bmatrix}
\ddots & \vdots & \vdots & \vdots & \iddots \\
\cdots & \kappa_{\displaystyle p_x',-1} - H' + V_{\displaystyle 0}  &  V_{\displaystyle -1} &  V_{\displaystyle -2} & \cdots \\
\cdots &  V_{\displaystyle 1} & \kappa_{\displaystyle p_x',0} - H' + V_{\displaystyle 0} & V_{\displaystyle -1} & \cdots \\
\cdots &  V_{\displaystyle 2} &  V_{\displaystyle 1} & \kappa_{\displaystyle p_x',1} - H' + V_{\displaystyle 0} & \cdots \\
\iddots & \vdots & \vdots & \vdots & \ddots
\end{bmatrix}
\cdot
\begin{bmatrix}
\vdots \\
D_{H',\displaystyle -1} \\
D_{H',\displaystyle 0} \\
D_{H',\displaystyle 1} \\
\vdots
\end{bmatrix}
=
\begin{bmatrix}
\vdots \\
0 \\
0 \\
0 \\
\vdots
\end{bmatrix}.
$}
\label{eqn31}
\end{equation}

\noindent Our goal is now accomplished: We have found the condition that makes Eq.~\eqref{eqn24} hold. The entire problem is reduced to this matrix-eigenproblem of infinite size. Here, for practical purposes, we truncate the infinite matrix in Eq.~\eqref{eqn31} to an \mbox{$N \times N$} matrix ($N$ being an arbitrary positive odd integer) and base our discussion on this truncated version. We start by noting that the matrix is Hermitian. This follows from the fact that the potential energy function, Eq.~\eqref{eqn12a}, is a real quantity: \mbox{$ V_{\ell} = V_{-\ell}^{*}$}. Secondly, we observe that the structure of this matrix eigenproblem gives rise to the term \textit{band index}. To see that, let us note: For a given certain $p_x'$, since the matrix is Hermitian and of size $N  \times N$, we expect it to have $N$ number of real eigenvalues, $H'$. And, also, we expect it to have $N$ number of eigenvectors (column-matrices) corresponding to each of these $H'$. We count these eigensolutions by introducing the non-negative integer, $b = \{ 0, 1, \dots, N-1 \}$, and call it the band index. However, to make this counting explicit, we have no choice but to alter our notation and rename the eigenvalue as: \mbox{$H' \rightarrow \scalebox{1}{$\epsilon$}_{p_x'}^{(b)}$}, and similarly, the elements of the corresponding eigenvector as: \mbox{$ D_{H',k} \rightarrow D_{p_x',k}^{(b)}$}. Such an alteration in the notation is necessitated by the intrinsic nature of the matrix eigenproblem. Within this naturally adapted notation, the subscript, $p_x'$, is linked to the same subscript of the $\kappa$ in the matrix, which, in turn, keeps track of the eigenvalue of $\hat{T}(a)$. And, the band index, $b$, labels the set of eigensolutions yielded by the matrix for a given $p_x'$, in an increasing fashion in energy. An additional index may be introduced to account for the possible degeneracies (band-crossing). Putting that possibility aside, what we have done is simply a notational bookkeeping to clarify the upcoming expressions. Now, let us revise Eq.~\eqref{eqn23} in light of this discussion (the following summations are understood to run from \mbox{$-(N-1)/2$} to \mbox{$(N-1)/2$}):

\begin{equation}
 \big| simul \big>_{p_x', b} = \sum_{k}^{} D_{p_x',k}^{(b)} \big|p_x' + 2 \pi \hbar k / a\big>,
 \label{eqn32}
\end{equation}

\noindent such that

\begin{subequations} \label{eqn33}
\begin{align}
    \hat{T}(a) \big| simul \big>_{p_x', b} &= \exp(-i \cdot \frac{p'_x}{\hbar} \cdot a) \big| simul \big>_{p_x', b}, \label{eqn33a} \\[10pt]
    \hat{H} \big| simul \big>_{p_x', b} &= \epsilon_{p'_x}^{(b)} \big| simul \big>_{p_x', b}. \label{eqn33b}
\end{align}
\end{subequations}

\noindent This is the main contribution of the paper: Equation~\eqref{eqn32} is the energy-eigenket corresponding to the \textbf{particular} energy-eigenvalue of $\scalebox{1}{$\epsilon$}_{p_x'}^{(b)}$. Here, \mbox{$p_x' \in \mathbb{R}$} and \mbox{$b = \{ 0, 1, \dots, N-1 \}$}. And, \mbox{$\big\{D_{p_x',k}^{(b)}\big\}$} and $\scalebox{1}{$\epsilon$}_{p_x'}^{(b)}$ are governed by the matrix-eigenproblem. With this concrete expression at hand, we can now carry it over to position space:

\begin{equation}
\begin{aligned}
\scalebox{1.0}{$\psi$}_{p_x',b} (x')  &\equiv  \big< x' \big|simul \big>_{p_x', b} \\
    &= \sum_{k}^{} \frac{D_{p_x',k}^{(b)}}{\sqrt{2 \pi \hbar}} \exp \Big( i \cdot \frac{p_x' + 2 \pi \hbar k / a}{\hbar} \cdot x' \Big) \\
    &= \frac{e^{i p'_x x' / \hbar} }{\sqrt{2 \pi \hbar} }  \cdot  \sum_{k}^{} D_{p_x',k}^{(b)} \exp \Big(i \cdot \frac{2 \pi k}{a} \cdot x'  \Big).
\end{aligned}
\label{eqn34}
\end{equation}

\noindent The second term in the last line is the familiar periodic modulation of the plane wave, and by writing Eq.~\eqref{eqn34} for $\psi_{p'_x,b} (x'+a)$ one obtains the well-known statement of the Bloch theorem. Notice how we obtained the wave-function without explicitly solving Schrödinger's wave-equation in position space. To proceed further, let us consider the probability density for the position-measurement:

\begin{equation}
\begin{aligned}
\big|\big|\scalebox{1.0}{$\psi$}_{p_x',b} (x') \big|\big|^{2}  &=  \scalebox{1.0}{$\psi$}_{p_x',b}^{*} (x') \scalebox{1.0}{$\psi$}_{p_x',b} (x')\\
    &= \sum_{j,k}^{}  \frac{D_{p_x',j}^{*(b)} D_{p_x',k}^{(b)}}{2 \pi \hbar}   \exp \Big(\frac{i 2 \pi (k-j) x'}{a} \Big).  \\  
\end{aligned}
\label{eqn35}
\end{equation}

\noindent We observe that the position probability density is a periodic function with a period of $a$. That is: \mbox{$||\scalebox{1}{$\psi$}_{p_x',b} (x' + a) ||^{2} = ||\scalebox{1}{$\psi$}_{p_x',b} (x')||^{2}$}. This observation underlies the reason \textit{why} a single electron in a periodic potential is inherently \textit{extended} through the entire lattice. We are going to clearly illustrate this in the upcoming section when we consider the Kronig-Penney model. But, before that, let us take a step further and consider also the momentum-measurement. From Eq.~\eqref{eqn32}, we observe that, on the simultaneous eigenket (i.e. the definite-energy state), labeled by a certain physical-momentum $p_x' \in \mathbb{R}$ and band index $b$, the momentum-measurement may yield only one of the discrete values of \mbox{$(p_x' + 2 \pi \hbar k /a)$} with a probability of $ ||D_{p_x',k}^{(b)}||^{2}$, for each integer $k$. This observation highlights that the physical-momentum serves merely as an index: It is not the momentum eigenvalue of the definite-energy state, but only a label that tracks the eigenvalue of the position-translation operator (see Appendix A). Furthermore, we also conclude that, unlike a delocalized free electron with a definite momentum, the Bloch electron is delocalized not only in the position space but also in the momentum space.

\section{Kronig-Penney Model}

In this section, we apply our analytical results to the Kronig-Penney model to illustrate our findings numerically. Let us start by introducing the model. We write the periodic potential as it is written in \hyperlink{references}{Ref. 9}: 

\begin{equation} \label{kronig}
V(x') = \sum_{j=-\infty}^{\infty} \frac{\hbar^2 \mathtt{P} }{m a} \cdot \delta(x' - j a).   
\end{equation}

\noindent Here, $\delta(\cdot)$ denotes the Dirac delta function, $\mathtt{P}$ sets the strength of each peak, and $m$ is the particle mass. Plugging this into Eq.~\eqref{eqn12b}, we obtain: \mbox{$V_{\ell}=\hbar^2 \mathtt{P} / m a^2$} for all $\ell$. Inserting this set of numbers into the truncated version of Eq.~\eqref{eqn31}, we get an eigenproblem for a real and symmetric matrix of size $N \times N$. Our task is to solve this simple matrix eigenproblem for each $p'_x$ value, in Python. To do so, we let \mbox{$\mathtt{P} = 6$}, \mbox{$ (-5\hbar \pi /a  \leq   p_x'  < 5\hbar \pi /a ) $} sampled into 4000 equally spaced pieces, and \mbox{$ (-3a  \leq   x'  < 3a ) $} sampled into 1000 equally spaced pieces. However, we emphasize that restricting the position eigenvalue to this interval, does not mean that we are considering a finite-chain. That is done only for the sake of numerical convenience. Let us now present the results for the $N=15$ case:

\begin{figure}[H]
    \centering
    \includegraphics[width=0.5\linewidth]{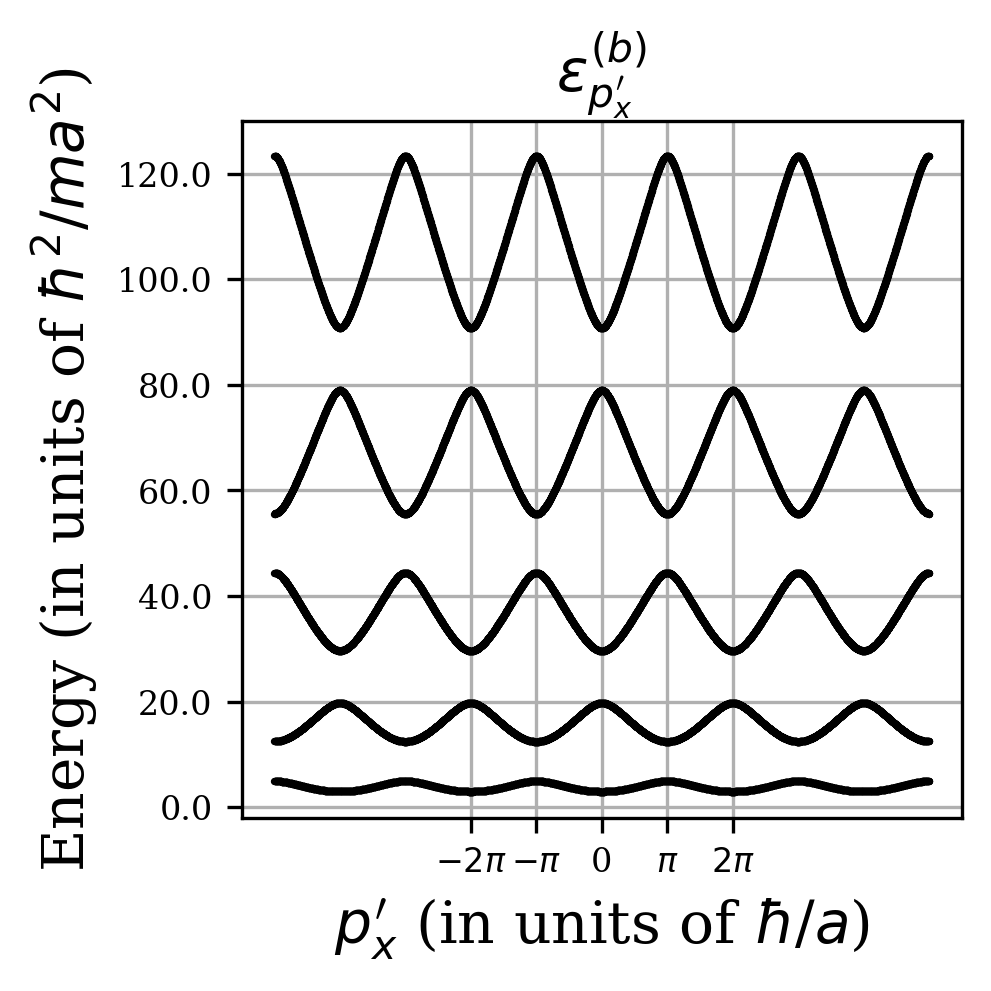}
    \caption{First five energy bands of the Kronig–Penney model. To eliminate the redundancy arising from the $p_x'$-periodicity, one may restrict the $x$-axis of the above plot to the interval $[-\pi\hbar/a,\, \pi\hbar/a)$. This interval is known as the \textit{first Brillouin zone}.}
    \label{dispersion}
\end{figure}

\noindent The plot in \Figref{dispersion} clearly demonstrates the $p_x'$-periodicity of the energy eigenvalues, and suggests the existence of forbidden intervals. These forbidden regions are referred to as \textit{band gaps}. Let us now pick a certain energy among these allowed values, and plot its state's position and momentum probability distributions:

\begin{figure}[H]
    \centering
    \begin{subfigure}[b]{0.48\linewidth}
        \centering
        \includegraphics[width=\linewidth]{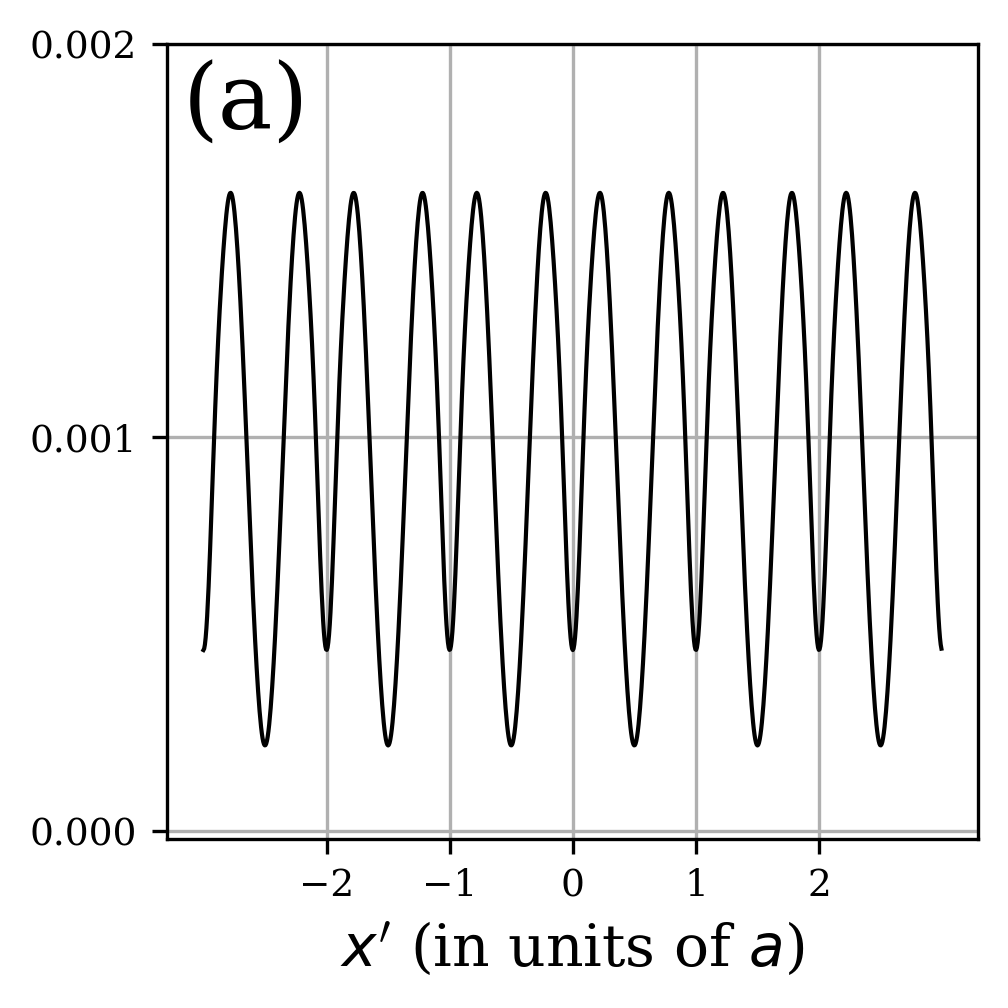}

    \end{subfigure}
    \hfill
    \begin{subfigure}[b]{0.45\linewidth}
        \centering
        \includegraphics[width=\linewidth]{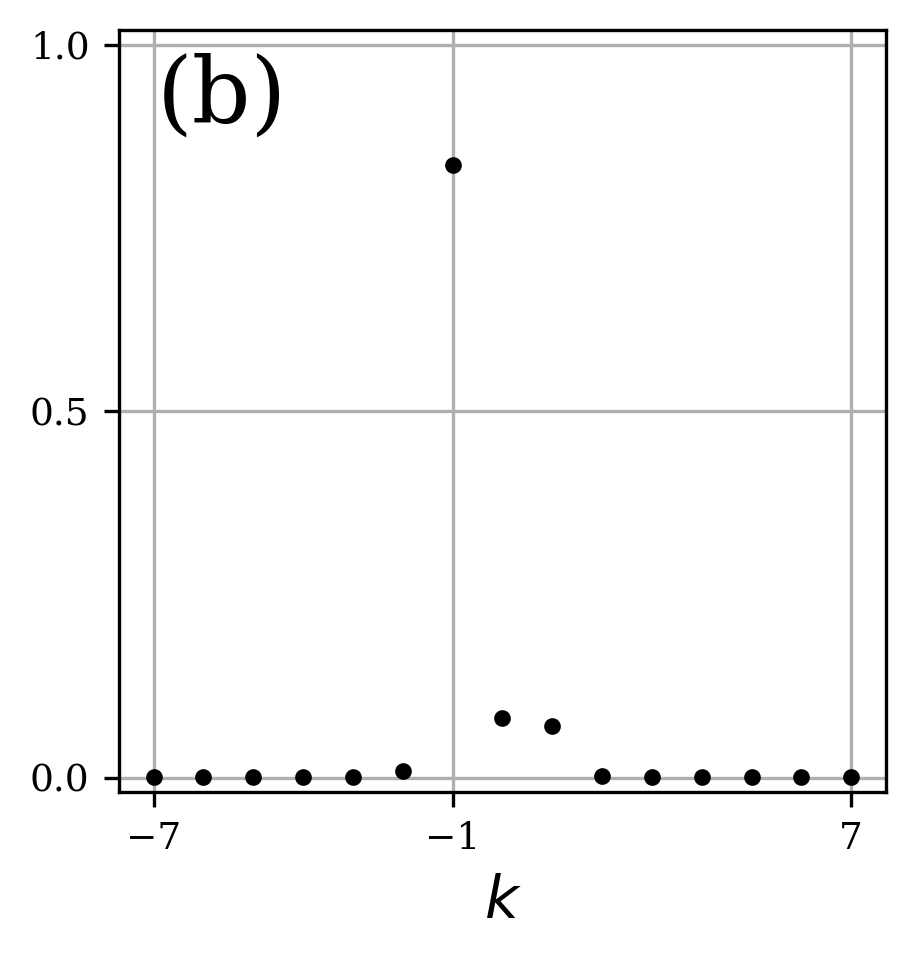}

    \end{subfigure}
    \caption{(a) Position and (b) momentum probability distributions for the definite-energy state occupying the second band ($b=1$) and labeled by the physical-momentum $p_x' = \pi \hbar / 2a$. The plotted distributions correspond to $  || \psi_{(\pi \hbar/2a),(1)}(x') || ^ {2}     $ and $  || D_{(\pi \hbar/2a),(k)}^{(1)} || ^ {2}  $, respectively.}   
    \label{fig:bothN}
\end{figure}

\noindent In \Figref{fig:bothN}, we present the position and momentum probability distributions for the definite-energy state inhabiting the second band (\mbox{$b=1$}) and labeled by the physical-momentum of \mbox{$p_x' = \pi \hbar/2a$}. Let us interpret the position probability density by quoting Bloch’s original motivation, as he described it, when Heisenberg assigned him the task: "When I started to think about it, I felt that the main problem was to explain how the electrons could sneak by all the ions in a metal so as to avoid a mean free path of the order of atomic distances."\textsuperscript{\hyperlink{references}{10}} In the context of the current model, by looking at the position probability density, we can see \textit{how} a single electron sneaks by all those Dirac delta potentials in Eq.~\eqref{kronig}. Contrary to classical intuition, the electron is neither scattered nor trapped; instead, it \textit{propagates} through the entire crystal. Let us now turn to the momentum probability distribution. The plot tells us that, on a definite-energy state labeled by the physical-momentum of \mbox{$p_x' = \pi \hbar/2a$} and $b=1$, the momentum-measurement is most likely to return the value of \mbox{$(\pi \hbar/2a + 2\pi\hbar \cdot (-1)/a =-3 \pi \hbar/2a)$}, or may seldom another value. Therefore, indeed, the label is not the momentum of the particle, but merely a marker for bookkeeping purposes. Recall how it did appear for the first time, as a subscript, in Eq.~\eqref{eqn23} (see Appendix A). We conclude this section by inviting the reader to numerically verify the position–momentum uncertainty relation, and to explore the free-particle case as well as other periodic potentials (e.g., the cosine potential) by appropriately modifying the Python code provided in the \hyperlink{references}{supplementary material}.

\section{Conclusion}

We conclude by noting that evaluating the sum for the first term in Eq.~\eqref{eqn30} yields precisely Kittel's \textit{central equation} in \hyperlink{references}{Ref. 6}, Eq. (8.41) in \hyperlink{references}{Ref. 4} and Eq. (3.37) in \hyperlink{references}{Ref. 5}:

\begin{equation}
\begin{aligned} 
\Bigg(  \frac{(p_x' + 2 \pi \hbar q / a)^2}{2m} - H' \bigg) D_{H',q} + \sum_{k=-\infty}^{\infty} V_{q-k}  D_{H',k} = 0.
\end{aligned}
\label{eqn36}
\end{equation}

\noindent Let us draw attention to a key variation: We have reproduced this classic result, the central equation, without relying on Schrödinger’s wave-mechanics approach. Instead, we employed a Heisenberg’s matrix-mechanics form of reasoning---the very framework developed by Felix Bloch’s own doctoral supervisor. Moreover, and most importantly, as a major advantage of this operator-based approach, in Eq.~\eqref{eqn32} we provide a concrete expression for the energy eigenket corresponding to a particular energy eigenvalue. This novel result enabled us to clearly illustrate the position and momentum probability distributions for a definite-energy state in \Figref{fig:bothN}, thereby providing students with an instructive demonstration. As a final pedagogic remark, the reader is invited to relate this treatment to Freericks's concluding question presented in \hyperlink{references}{Ref. 11}: "Is the wavefunction-based approach really the best way to teach quantum mechanics? Or, can we learn something new about alternative ways to proceed based on a matrix mechanics form of reasoning?" In the same spirit, this work presents a novel, operator-based, derivation of Bloch's theorem, offering a clear alternative to the conventional differential-equation-based approaches, while also providing valuable insights for students of quantum mechanics and condensed matter physics. This treatment may be considered as an extension of Park's problem 5.11 in \hyperlink{references}{Ref. 12}, which is also an operator-based approach.

\section*{Appendix A: \\ The Convenient Label and Crystal Momentum}

\renewcommand{\theequation}{A\arabic{equation}}
\setcounter{equation}{0}

A reader familiar with conventional treatments of Bloch’s theorem, may notice a difference in the labeling (i.e. the use of $p'_x$ instead of \textit{crystal momentum}$/\hbar$). It is instructive at this point to examine the source of this apparent variation. For this purpose, let us refer to Ashcroft\&Mermin\textsuperscript{\hyperlink{references}{4}} and investigate the origin of their choice of label. The distinction begins to emerge at the outset on p. 134. There, they define the position-translation operator in Eq. (8.7). And then, by inspection, they write its eigenvalue in Eq. (8.19):

\begin{equation}
    c(\boldsymbol{R}) = e^{i \boldsymbol{k} \cdot \boldsymbol{R}}. 
\end{equation}

\noindent This is the first instance in which the \textit{wave vector}, $\boldsymbol{k}$, appears in this context. It then serves as a convenient label for the simultaneous eigenstate and, on p. 139, $\hbar \boldsymbol{k}$ is identified as the \textit{crystal momentum}. Whereas in this operator-based treatment, since we derive the position-translation operator from first principles, using the fundamental commutation relation $[\hat{x}, \hat{p}_x] = i \hbar \cdot \hat{I}$ and the Hadamard lemma, we prove that its eigenvalue is \mbox{$\exp (-i p_x'  a / \hbar )$}, and therefore the convenient label in this treatment turns out to be the momentum itself, \mbox{$p_x' \in \mathbb{R}$}. Compare Eqs.~\eqref{eqn33} with Eqs. (8.12) of Ashcroft\&Mermin. Let us reiterate the crucial point: The label emerges solely as an artifact of the method of simultaneous diagonalization, and it should not be interpreted as the momentum of the definite-energy state. This aspect is illustrated in \Figref{fig:bothN} for the Kronig-Penney model.

In this framework, the term \textit{crystal momentum} can be introduced as a possible outcome of the momentum measurement. As discussed at the end of Sec. III, the momentum measurement can collapse the state only onto a definite-momentum eigenstate whose momentum differs from the label by $2\pi \hbar k / a$, and not onto an arbitrary momentum eigenstate. These evenly spaced discrete outcomes can be identified as the \textit{crystal momenta}.

\section*{Appendix B:\\ Operator Identities}

\renewcommand{\theequation}{B\arabic{equation}}
\setcounter{equation}{0}

\subsection{Hadamard Lemma}

Recall the Taylor formula:

\begin{equation}
\begin{aligned}
    \hat{f}(\lambda) 
    &= \hat{f}( 0) + \lambda \cdot \left.\frac{d\hat{f}(\lambda)}{d \lambda} \right|_{\lambda=0} + \frac{\lambda^{2}}{2!} \cdot \left.\frac{d^{2}\hat{f}(\lambda)}{d \lambda^{2}} \right|_{\lambda=0} + \dots
\end{aligned}
\label{eqnA1}
\end{equation}

\noindent Here, $\lambda$ is a continuous variable and the function $\hat{f}(\lambda)$ is an operator. To derive Eq.~\eqref{opidentityb}, consider the following function:

\begin{equation}
    \hat{f}(\lambda) = e^{\lambda \hat{A}} \cdot \hat{C} \cdot e^{-\lambda \hat{A}}.
    \label{eqnA4}
\end{equation}

\noindent By differentiating this using the product rule, and plugging it into Eq.~\eqref{eqnA1}; the pattern in Eq.~\eqref{opidentityb} can be established when evaluated at \mbox{$\lambda=1$}.

\subsection{Braiding Identity}

Consider the product $e^{\xi \hat{x}} \cdot e^{\zeta \hat{p}_x} \cdot e^{-\xi \hat{x}}$ and plug in the operator-series for the exponential in the middle:

\begin{equation} \label{bra}
    e^{\xi \hat{x}} \cdot e^{\zeta \hat{p}_x} \cdot e^{-\xi \hat{x}} = \sum_{n=0}^{\infty} \frac{e^{\xi \hat{x}} \cdot      (\zeta \hat{p}_x)^n         \cdot e^{-\xi \hat{x}}    }{n!}.
\end{equation}

\noindent Inserting $(n-1)$ number of $e^{- \xi \hat{x}} \cdot e^{ \xi \hat{x}} = \hat{I} $ into the proper places in the numerator of the right-hand-side yields: \mbox{ $ \exp (\xi \hat{x}) \cdot (\zeta \hat{p}_x)^{n} \cdot \exp (- \xi \hat{x}) = ( \exp (\xi \hat{x}) \cdot \zeta \hat{p}_x \cdot \exp (- \xi \hat{x}) )^{n} $}. Therefore, we have:

\begin{equation}
    e^{\xi \hat{x}} \cdot e^{\zeta \hat{p}_x} \cdot e^{-\xi \hat{x}} = \exp (e^{\xi \hat{x}} \cdot \zeta \hat{p}_x  \cdot e^{-\xi \hat{x}} ).
\end{equation}

\noindent This gives the first line of Eq.~\eqref{eqn10} and is known as the braiding identity.

\section*{Supplementary Material}
The Python code for the Kronig–Penney model (Sec. IV) is provided as a .txt file and is available at: \href{https://doi.org/10.60893/figshare.ajp.c.8467149}{https://doi.org/10.60893/figshare.ajp.c.8467149}.

\section*{Acknowledgments}
I am deeply grateful to Professor James K. Freericks for providing multiple rounds of comprehensive feedback and for his generous long-distance mentorship. I am also thankful to Professor Jim Napolitano and Professor Nejat Bulut for their encouragement and support. I further thank the anonymous referees for their insightful suggestions.

\vspace{0.5em}

\noindent I declare no conflicts of interest.

\hfill

\hypertarget{references}{}

\begingroup
\fontsize{8.5pt}{10pt}\selectfont  

\noindent $^{a)}$Email: celalsirin@iyte.edu.tr, ORCID: \href{https://orcid.org/0009-0005-1863-6773}{0009-0005-1863-6773}.

\noindent $^{1}$Felix Bloch, "Über die Quantenmechanik der Elektronen in Kristallgittern," \href{https://doi.org/10.1007/BF01339455}{Z. Physik} \textbf{52}, 555--600 (1929).

\noindent $^{2}$R. De L. Kronig and William George Penney, "Quantum mechanics of electrons in crystal lattices," \href{https://doi.org/10.1098/rspa.1931.0019}{Proc. R. Soc. Lond.} A \textbf{130}: 499–513 (1931). 

\noindent $^{3}$M. S. P. Eastham, \textit{The Spectral Theory of Periodic Differential Equations}, (Scottish Academic Press, 1973), Chapter 6.

\noindent $^{4}$Neil W. Ashcroft and N. David Mermin, \textit{Solid State Physics}, (Harcourt College Publishers, 1976), p. 138.

\noindent $^{5}$Efthimios Kaxiras, \textit{Atomic and Electronic Structure of Solids}, (Cambridge University Press, 2003), p. 93.

\noindent $^{6}$Charles Kittel, \textit{Introduction to Solid State Physics}, 8th Ed. (John Wiley \& Sons, 2005), p. 172.

\noindent $^{7}$Eugen Merzbacher, \textit{Quantum Mechanics}, 2nd Ed. (Wiley International Edition, John Wiley \& Sons, 1970), pp. 157–158.

\noindent $^{8}$J. J. Sakurai and Jim Napolitano, \textit{Modern Quantum Mechanics}, 3rd Ed. (Cambridge University Press, 2021), Chapter 1.

\noindent $^{9}$Surjit Singh, "Kronig--Penney model in reciprocal lattice space," \href{https://doi.org/10.1119/1.13321}{Am. J. Phys.} \textbf{51} (2), 179 (1983). 

\noindent $^{10}$National Academy of Sciences, \textit{Biographical Memoirs: Volume 64} (Washington, DC: The National Academies Press, 1994), p. 41.

\noindent $^{11}$James K. Freericks, "Why is it so difficult to generalize Heisenberg’s matrix mechanics from the harmonic oscillator to other exactly solvable problems?" \href{https://doi.org/10.1140/epjs/s11734-025-01764-z}{Eur. Phys. J. Spec. Top.} (2025).

\noindent $^{12}$David Park, \textit{Introduction to the Quantum Theory}, 2nd Ed. (McGraw-Hill, 1974), p. 149.

\endgroup

\end{document}